# Automated Registration of 3D Neurovascular Territory Atlas to 2D DSA for Targeted Quantitative Angiography Analysis

George Dimopoulos [1,2], Sabrina De Los Angeles Reverol Parra[1,2], Parmita Mondal[1,2], Michael Udin[1,2], Kyle Williams[1,2], Parisa Naghdi[1,2], Ahmad Rahmatpour[1,2], Swetadri Vasan Setlur Nagesh[1,2], Mohammad Mahdi Shiraz Bhurwani[3], Jason Davies[1,4] and Ciprian N. Ionita [1,2]

[1]Canon Stroke and Vascular Research Center, Buffalo, NY, USA

[2]Department of Biomedical Engineering, State University of New York at Buffalo, Buffalo, NY, USA,

[3]QAS.AI Inc.

[4]Department of Neurosurgery, State University of New York at Buffalo, Buffalo, NY, USA

## ABSTRACT

**Background:** Subarachnoid hemorrhage (SAH), typically due to intracranial aneurysms, demands precise imaging for effective treatment. Digital Subtraction Angiography (DSA), despite being the gold standard, broadly visualizes cerebral blood flow, potentially masking key details in areas such as the Middle Cerebral Artery (MCA) or Anterior Cerebral Artery (ACA). This study introduces an approach integrating a 3D vascular atlas from the Brain Arterial Vascular Model project with 2D DSA images to allow targeted quantitative analysis in these crucial regions, thus enhancing diagnostic accuracy during interventions.

**Methods:** Initially, DSA data was examined to ascertain the injection site—left anterior, right anterior, or posterior. Following this, the appropriate viewing angle, lateral or anteroposterior, was determined to align accurately with the 3D vascular atlas. Utilizing this atlas, regions corresponding to the areas indicated as perfused were selected. A cone beam projection was then performed on the atlas, employing its geometrical information to accurately simulate the 2D projection of these selected 3D regions. Concurrently, a mask representing the perfused areas was created from the DSA sequence. This mask facilitated the initial coarse alignment of the projected 3D atlas to the DSA perfused territory through affine transformations. Further refinement of this alignment was achieved using deformable registration techniques, ensuring a precise overlay with the DSA's perfused territories. The performance of each overlay was measured using the Structural Similarity Index Measure (SSIM).

**Results:** The coregistration process revealed that affine transformations alone were insufficient for accurate alignment due to the variability in brain anatomy across individuals. Consequently, deformable registration was essential to achieve precise overlays of the 3D atlas projections with the 2D DSA perfused areas. This more accurate overlay was able to be segmented into various arterial territories, taking on average 1 to 3 minutes from loading the DSA to coregistering, overlaying, and segmenting the modalities. This approach enabled the extraction of targeted quantitative angiography parameters, essential for detailed vascular assessment in subarachnoid hemorrhage cases.

**Conclusion:** The integration of 3D atlas registration with 2D DSA projections facilitates a more precise and targeted diagnostic process for SAH during critical interventions. This image processing strategy enhances the visualization of affected arterial territories, potentially improving the accuracy of diagnostics and supporting better-informed clinical decisions at the time of intervention.

## Summary

In this study, we investigate the automated segmentation of angiographic images into arterial territories from patients with Subarachnoid Hemorrhage (SAH). We utilize advanced registration techniques to map a 3D model atlas to real patient data, generating a segmented and annotated overlay to yield targeted quantitative angiographic parameters. The integration of 3D atlas registration with 2D DSA projections facilitates a more precise and targeted diagnostic assessment of SAH during interventional procedures. This image processing strategy enhances the visualization of affected arterial territories, potentially improving the accuracy of diagnostics and supporting better-informed clinical decisions at the time of intervention.

## 1. INTRODUCTION

Intracranial aneurysms (IAs) are a set of severe neurological conditions presenting as dilations in the arteries and pose significant risks to the health of the patient. One form of IA, which occurs when the ruptured artery bleeds in the subarachnoid space is subarachnoid hemorrhage (SAH) - a form of stroke that accounts for around a quarter of cerebrovascular deaths despite occurring in only 6% of the population. [1-2] Morbidity and mortality rates remain dangerously high, with approximately 65% of patients dying from the initial bleed or its complications, and half of all survivors left significantly disabled. [3] The pathophysiology of SAH involves complex interactions between the leaking blood and cerebral tissues, potentially triggering vasospasm, delayed cerebral ischemia (DCI), and a cascade of inflammatory responses that significantly affect patient outcomes. [4-6]

Currently, the endovascular treatment of SAH during interventions primarily relies on 2D Digital Subtraction Angiography (DSA) methods, which lack the capability to provide detailed insights into vascular territories during critical interventions. Recent advancements in quantitative angiographic methods and imaging technologies promise to enhance the quality and effectiveness of these interventions. For instance, quantitative angiography parametric imaging (API) has been explored for its ability to provide detailed assessments of tissue perfusion and vascular disease severity. [7-10]

This study introduces an innovative approach that combines 3D atlas registration with 2D DSA projections to perform targeted quantitative angiography. By integrating detailed 3D vascular territories into 2D imaging used during interventions, this method aims to overcome the limitations of traditional quantitative angiography by allowing precise localization and assessment of specific vascular territories affected by SAH. This approach is hypothesized to enhance diagnostic accuracy and enable a more nuanced understanding of the vascular changes associated with SAH, potentially improving prognosis and patient outcomes by facilitating timely and targeted interventions.

## 2. MATERIALS AND METHODS

Patients treated at the Gates Vascular Institute were selected for this study following Institutional Review Board approval. A physician verified all imaging samples to confirm the presence of SAH. High-quality angiographic images were selected based on the visibility of contrast flow, excluding frames where the carotid artery was visible to leave us with only the relevant information of the brain structure. Samples with patient or equipment motion artifacts or insufficient frame sequences to track contrast dispersion were also excluded.at

First, expert users manually labeled each sequence based on the site of the injection—left anterior, right anterior, or posterior. For posterior injections, it was noted that both vertebral arteries converge into the basilar artery, perfusing the entire posterior brain region, thus requiring no further subdivision. Furthermore, since the data was acquired using a biplane angiographic system, the projection view of the DSA data, either lateral or posterior, was identified to ensure the rapid and accurate alignment of the vascular atlas with the 2D angiographic images. Based on the identified labels for the injection site and view, corresponding perfused regions were selected from the 3D vascular atlas using a lookup table. We utilized an atlas from the Brain Arterial Vascular Model project, where each arterial region is assigned a specific number [11]. Using the geometry information from the DSA's DICOM header, a cone beam projection of the selected 3D atlas regions was simulated using the ASTRA toolbox. This simulation of 2D projections of the 3D regions was essential for aligning the atlas with the actual DSA images.

In parallel with atlas projection creation, the DSA sequences were first temporally-averaged, then thresholded to create masks representing the perfused areas. This was followed by removing small, connected components that did not meet a minimum size criterion, effectively filtering out image artifacts and noise. The resulting binary image was further refined using morphological operations such as grey erosion to clean up the mask edges and binary fill holes to ensure the mask accurately represented the perfused areas. These steps resulted in a clean and precise mask highlighting the perfused regions within the DSA images, which was crucial for subsequent alignment and quantitative analysis.

Using Python's SimpleITK library, an initial affine transformation was applied to co-register the projected 3D atlas with the DSA mask, establishing a coarse alignment that broadly matched the geometrical configuration of the atlas with the angiographic data. The affine transformation parameters were iteratively determined using a mutual information cost function, with the number of resolutions set to 20, and a quasi-Newton LBFGs optimizer. The maximum step length for the optimizer was set to 2, and automatic scale estimation was enabled to adjust the transformation scale dynamically. This initial alignment provided a foundational geometric match, which was crucial for the subsequent fine-tuning steps using deformable registration techniques.

Following the initial affine transformation, the alignment was further refined using deformable registration techniques to account for individual anatomical differences and achieve a precise overlay of the projected atlas onto the DSA images. This refinement utilized B-spline transformations implemented through the SimpleITK library. The B-spline registration parameters were iteratively determined using a mutual information cost function, with the number of resolutions set to 16, and a quasi-Newton LBFGs optimizer with a maximum step length of 2. This deformable registration step ensured a highly accurate alignment of the atlas to the DSA images, enabling detailed and precise mapping of vascular territories. Following the B-spline registration, performance of each overlay was quantified using the SSIM, with reference to the processed DSA data.

## 3. RESULTS

After achieving refined coregistration, each arterial region from the 3D vascular atlas was individually processed to overlay onto the DSA images. The transformation parameters for both affine and B-spline registrations were saved for each arterial region, facilitating future quantitative angiography analysis. This process enabled precise localization and detailed assessment of blood flow dynamics within specific vascular territories, essential for understanding perfusion characteristics and evaluating vascular health in patients with subarachnoid hemorrhage. These specific vascular territories were extracted from the overlay, which took arterial territory information from the 3D atlas that it was based on, plotting the locations of which regions the processed DSA possessed.

2,247 DSAs were analyzed. Following automatic registration and segmentation, an average SSIM of $0.69 \pm 0.27$ was observed. Exceptional examples of registration were observed in higher proportions in anteroposterior views compared to lateral views. Figure 1 shows a quality overlay where the DSA was effectively processed, and the registration fits the image to a high degree. A large variance was observed in the SSIM results, with many of the values being concentrated close to 1, and comparatively fewer values being closer to 0, thus skewing the mean performance to be much less than the median, which was measured to be 0.81. This produced a strongly left skewed histogram of the results, as seen in Figure 2. Overall, our algorithm was able to process a DSA effectively and generate an arterial atlas overlay to produce a segmented map.

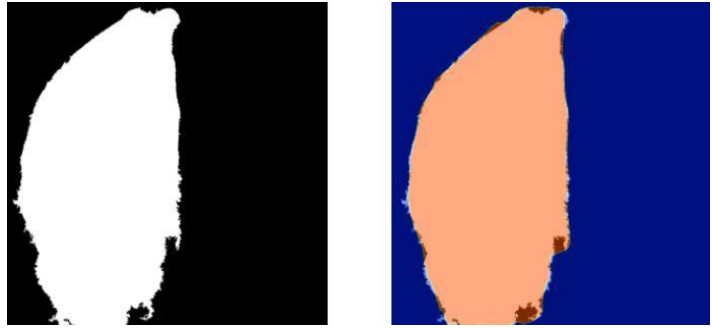

Figure 1: Left: The Processed DSA data; Right: B-Spline Registration. The DSA is processed, and the registration correctly overlays the shape of the brain, ignoring irrelevant information produced by noise in the image, yielding an SSIM of 0.96.

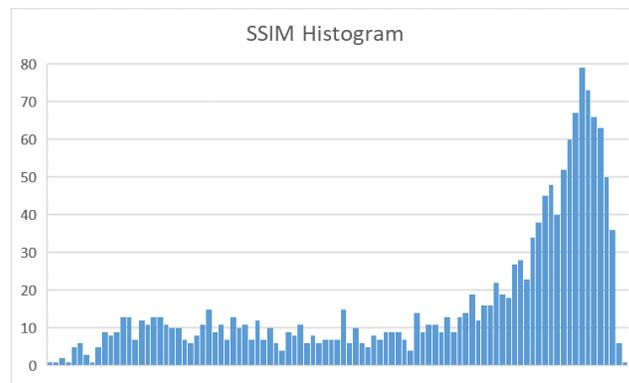

Figure 2: SSIM Histogram; shows the distribution of results. A bin width of 0.01 was used, with the vertical axis representing the total number (n) of cases that fall in the range.

## 4. CONCLUSIONS

This study demonstrates a novel approach for enhancing the precision of SAH diagnosis and intervention through the integration of a 3D vascular atlas with 2D DSA. By systematically labeling DSA sequences and utilizing advanced cone beam projection techniques, we achieved accurate alignment of the 3D atlas with 2D DSA images using affine and B-spline coregistration. The ability to isolate and analyze individual vascular territories is crucial for understanding the complex hemodynamics associated with SAH and potentially improving patient outcomes. Although quantitative analysis was not included in this study, the coregistration parameters saved during this process provide a robust foundation for future quantitative assessments. Our approach underscores the potential of combining advanced imaging techniques with atlas-based registration to enhance the diagnostic and therapeutic strategies for cerebrovascular diseases.

### Speaker Biography

George Dimopoulos is an undergraduate research assistant at the Canon Stroke and Vascular Research Center. He is pursuing his bachelor's in biomedical engineering from the University at Buffalo. Under the guidance of Dr. Ionita, he investigates how medical data can be understood and used. He would be delighted to further his education by attending medical school with the goal of practicing as a neuroradiologist. In addition to his studies, George is an active composer and trombonist, performing in orchestras and writing his own works.